\documentclass[prb,twocolumn,showpacs,superscriptaddress,preprintnumbers,amssymb]{revtex4}
\usepackage{graphicx}% Include figure files
\usepackage{dcolumn}% Align table columns on decimal point
\usepackage{bm}% bold math
\usepackage{wasysym}
\newcommand{\beq}{\begin{equation}}
\newcommand{\eeq}{\end{equation}}
\newcommand{\beqn}{\begin{eqnarray}}
\newcommand{\eeqn}{\end{eqnarray}}

\begin{document}

\title{Quantum Phase Transitions around the Staggered Valence Bond Solid}
\author{Cenke Xu}
\affiliation{Department of Physics, University of California,
Santa Barbara, CA 93106}
\author{Leon Balents}
\affiliation{Department of Physics, University of California,
Santa Barbara, CA 93106} \affiliation{Kavli Institute for
Theoretical Physics, University of California, Santa Barbara, CA,
93106}

\date{\today}

\begin{abstract}

Motivated by recent numerical results, we study the quantum phase
transitions between $Z_2$ spin liquid, N\'{e}el ordered, and
various valence bond solid (VBS) states on the honeycomb and
square lattices, with emphasis on the staggered VBS. In contrast
to the well-understood columnar VBS order, the staggered VBS is
not described by an XY order parameter with $Z_N$ anisotropy close
to these quantum phase transitions. Instead, we demonstrate that
on the honeycomb lattice, the staggered VBS is more appropriately
described as an O(3) or CP(2) order parameter with cubic
anisotropy, while on the square lattice it is described by an O(4)
or CP(3) order parameter.
%, plus lattice depended anisotropies.

\end{abstract}
 \maketitle

\section{introduction} \label{introduction}

Exotic criticality, in particular transitions which violate the
Landau rules for continuous phase transitions, is now believed to
be possible and perhaps even prevalent at quantum critical points.
The best-studied example of such phenomena is the ``Deconfined
Quantum Critical Point'' (DQCP) between the columnar valence bond
solid ($c-$VBS, Fig.~\ref{honeyvison1}$a$,
Fig.~\ref{honeyvison2}$b$) and N\'{e}el ordered antiferromagnet,
in simple unfrustrated geometries such as the square and honeycomb
lattices.  An intuitive understanding of this transition is
available through the topological defects of these phases.  Coming
from the antiferromagnetic state, one can regard the transition as
the proliferation of the skyrmion of the N\'{e}el order parameter
\cite{senthil2004a,senthil2004}, which destroys the N\'eel order.
It simultaneously creates valence bond solid order, because the
skyrmion carries the same quantum number as $c-$VBS
\cite{haldanemonopole,sachdev1990} on both the honeycomb and the
square lattices.  In this picture, we can view the skyrmion as a
boson, and the $c-$VBS order simply corresponds to the superfluid
phase of the skyrmion boson.  Although this skyrmion boson is not
precisely a conserved particle, it is expected that at the
deconfined quantum critical point, the skyrmions is fully
conserved in the long wavelength limit.  Hence it is reasonable to
describe the $c-$VBS with an XY order parameter, whose
anisotropies become irrelevant at the critical point.  This is a
specific {\sl embedding} of the discrete $c-$VBS order parameter
in a larger XY (U(1)) order parameter space. A consequence of the
irrelevance of anisotropy in this XY space is that the $c-$VBS
order is ``unified'' with plaquette VBS order, such that both
types of states are nearly degenerate near the critical point.  A
similar picture can be applied to the transition between the $Z_2$
spin liquid phase and the $c-$VBS, where the XY order parameter is
the vison field in the $Z_2$ spin liquid. This picture has been
confirmed with quantum Monte Carlo simulation on the $J-Q$ spin
models with multi-spin interactions, and it was clearly shown that
the U(1) symmetry of the $c-$VBS order parameter is fully restored
at the critical point between N\'{e}el and $c-$VBS
\cite{sandvik2009}.

\begin{figure}
\includegraphics[width=3.4 in]{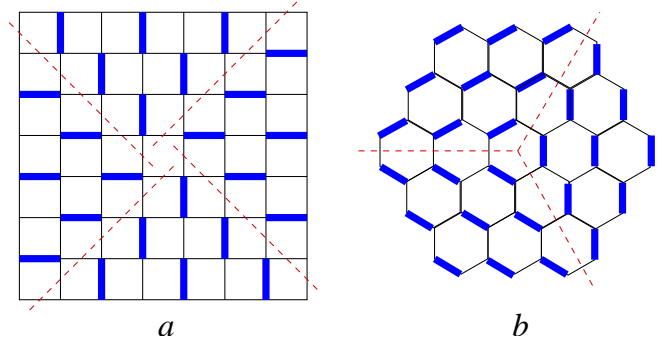}
\caption{The $Z_4$ vortex and $Z_3$ vortex of $s-$VBS on the
square and honeycomb lattice, the vortex core is featureless.}
\label{sVBSdefect}
\end{figure}

The staggered VBS ($s-$VBS) is another very natural pattern on
both the honeycomb and the square lattices
(Fig.~\ref{honeyvison1}$b$, Fig.~\ref{honeyvison2}$c$), and quite
distinct from the $c-$VBS. However, so far there has been no
theory describing the transition from magnetic ordered phases or
spin liquid to the $s-$VBS. The difference between the $c-$VBS and
$s-$VBS is hinted at by their vortices. It was noticed that a
vortex of $c-$VBS always carries an unpaired spinon, thus the
proliferation of $c-$VBS vortices will lead to magnetic order.
However, in Fig.~\ref{sVBSdefect} it is clearly shown that the
vortex of $s-$VBS is completely featureless, hence the transition
into magnetic order cannot be driven by these vortices.  Therefore
a completely different theory is needed to describe the $s-$VBS
and its quantum phase transitions.

It was proposed that for spin-1/2 systems on the honeycomb
lattice, the quantum fluctuations tend to melt the incommensurate
spin spiral order, and induce a $s-$VBS \cite{paramekantisvbs}.
Also, phase diagrams involving the $s-$VBS have been proposed in
many recent numerical works. For instance, exact diagonalization
of the $J_1-J_2$ Heisenberg model on the honeycomb lattice found a
gapped liquid phase between the N\'{e}el order and $s-$VBS
\cite{claire2001}, and a similar phase diagram of this model
including the liquid phase and the $s-$VBS was later confirmed
with variational methods \cite{clark2010}. Strong tendency towards
the $s-$VBS was also verified by functional renormalization group
studies on the same model \cite{thomalehoney}. On the square
lattice, exact diagonalization of the $J-Q$ model
\cite{claire2004} suggests that the N\'{e}el order can have direct
transition into both $c-$VBS and $s-$VBS. QMC of a modified $J-Q$
model on the square lattice also revealed a direct transition
between N\'{e}el order and $s-$VBS (although it was found to be
first order in this case) \cite{sandvik2009a}.

These numerical results strongly suggest that the $s-$VBS is a
very important competing order on the honeycomb lattice and square
lattice. In this paper we develop a theory of phase transitions
to/from the $s-$VBS.  Put simply, the main result of our analysis
is that the appropriate embedding for the $s-$VBS order parameter
is into a vector (O(n)) or complex projective (CP(n)) space, both
very different from the XY order parameter of the $c-$VBS case. In
the main text, we propose effective field theories for several
quantum critical points neighboring the $s-$VBS phase, which
expose these embeddings.  As in the original proposal for the
DQCP, these are strongly coupled field theories in $2+1$
dimensions, so some of their properties cannot be established
conclusively from analytics (we discuss in the main text what can
be inferred from the renormalization group literature).  However,
a robust prediction of the proposed field theories is that, like
in the N\'eel--$c-$VBS problem, the embedding into a larger order
parameter space unifies $s-$VBS order with other competing orders,
all of which should be present in low energy fluctuations in the
vicinity of the quantum critical points.  The nature of the
competing order(s) is determined from our analysis.  For instance,
at the transition between the $s-$VBS and a $Z_2$ spin liquid, the
competing order is of four-sublattice plaquette VBS type
(Fig.~\ref{honeyvison1}c).  Other cases are described in the main
text.  The presence of such competing orders at low energies is
testable in numerics or experiment.

The remainder of the paper is organized as follows.  In sections
\ref{z2honeycomb} and \ref{z2square}, we discuss the theory
describing the phase transition between the $Z_2$ spin liquid and
various VBS orders, including the $s-$VBS. In section
\ref{ordervbs}, this theory is extended to the transition between
the (easy-plane) N\'{e}el order and $s-$VBS.  We conclude with
some general remarks in Sec.~\ref{sec:summary-outlook}.

\section{Odd $Z_2$ gauge field on the honeycomb lattice} \label{z2honeycomb}

In this section we begin by considering the simplest candidate for
the fully gapped spin liquid state observed in recent numerics on
the honeycomb lattice \cite{meng,claire2001,clark2010}, which is
the $Z_2$ spin liquid.  We will study the transition between the
$Z_2$ liquid phase and different types of VBS orders using an
effective $Z_2$ gauge theory on the honeycomb lattice, and
eventually in this way connect to the $s-$VBS state.  To perform
the analysis, we will work with a lattice $Z_2$ gauge theory,
which is particularly convenient.  One may wonder whether this is
a sufficient and general starting point, correct for all possible
$Z_2$ spin liquid states, since many distinct such states are
possible using the projective symmetry group analysis of candidate
wavefunctions (see, e.g. Wen's analysis of the square lattice
\cite{wen2002a}). The answer is, we believe, yes, since our
analysis of the lattice gauge theory in fact rests only on three
key assumptions, which are true for all fully gapped symmetric
$Z_2$ spin liquids: (1) the $Z_2$ state supports ``visons'' ($Z_2$
vortices), which have a mutual statistics angle of $\pi$ with
respect to elementary spin-$1/2$ spins; (2) the $Z_2$ state
preserves all the lattice symmetries; and (3) there is a gap to
all excitations in the spin liquid state.  Thus we believe the
results of the following analysis are generally true for
transitions from arbitrary gapped $Z_2$ spin liquids to VBS
states.

To understand the meaning of the Ising gauge theory, consider
constructing VBS ordered states as the limit of ``hard dimers'',
in which precisely one dimer (spin singlet) is attached to each
site.  The dimer constraint is translated into the Gauss law
constraint in the gauge field language: $\vec{\nabla} \cdot
\vec{e} = \eta_i $, and $\eta_i = \pm 1$ on two different
sublattices. Now if the U(1) gauge symmetry is broken down to
$Z_2$, we need to introduce a $Z_2$ electric field $\sigma^x_{ij}
= (-1)^{n_{ij}}$ on every link ($n_{ij} = 0, 1$ denotes the
presence and absence of dimer), and the gauge constraint becomes
\beqn \prod_{\mathrm{links} \ \mathrm{round} \ \mathrm{site} \
  i} \sigma^x_{ij} = -1. \label{constraint}\eeqn With this $Z_2$ gauge
constraint, we can write down the simplest $Z_2$ gauge theory on
the honeycomb lattice as follows: \beqn H = \sum_{\hexagon} - K
\prod_{\mathrm{links} \ \mathrm{in} \ \hexagon}^6 \sigma^z_{ij} -
\sum_{i,j} h \sigma^x_{ij} + \cdots. \label{z2gauge}\eeqn The
first term is a sum of the ring product of the $Z_2$ gauge field
$\sigma^z_{ij}$ in every hexagon, and the second term is a $Z_2$
``string tension". The ellipses include other interaction terms
between $Z_2$ electric field.

\begin{figure}
\includegraphics[width=3.2 in]{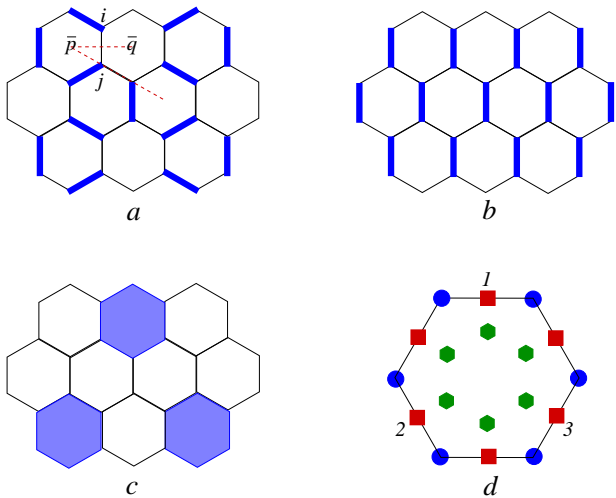}
\caption{({\it a}), $c-$VBS order. $\bar{p}$ and $\bar{q}$ are the
dual triangular lattice sites. We consider the nearest and 2nd
neighbor hopping for vison (vortex). ({\it b}), the $s-$VBS
pattern, realized when $h/8 < J < h$ in the dual Ising Hamiltonian
Eq.~\ref{jh}. ({\it c}), the four sublattice plaquette order,
realized when $w > 0$ in Eq.~\ref{cubic}. ({\it d}), the vison
(vortex) Brillouin zone. For weak 2nd neighbor vison (vortex)
hopping, the minima of band structure are located at the corner of
the BZ (circles); with intermediate 2nd neighbor hopping, there
are three inequivalent minima located at the center of the edges
of BZ (square); There are six inequivalent incommensurate minima
with strong 2nd neighbor hopping (hexagon). } \label{honeyvison1}
\end{figure}

When the $K$ term dominates everything else in Eq.~\ref{z2gauge},
the system is in the deconfined phase of the $Z_2$ gauge theory,
with topological degeneracy. When $h$ or other interaction terms
between $\sigma^x$ dominate $K$, the system enters the confined
phase. In order to analyze the confined phase, it is conventient
to go to the dual picture of the $Z_2$ gauge theory. Dual
variables $\tau^z$ and $\tau^x$ are defined on the dual lattice
sites $\bar{m}$, which are located at the center of the hexagons
(Fig.~\ref{honeyvison1}$a$): \beqn && \sigma^x_{ij} = -
\tau^z_{\bar{p}}\tau^z_{\bar{q}}, \ \ \bar{p} \ \mathrm{and} \
\bar{q} \ \mathrm{share} \ \mathrm{link} \ ij, \cr\cr &&
\prod_{\mathrm{links} \ \mathrm{around} \ \bar{p}}^6 \sigma^z_{ij}
= \tau^x_{\bar{p}}. \label{dual}\eeqn Introduction of
$\tau^z_{\bar{i}}$ automatically solves the odd $Z_2$ gauge
constraint Eq.~\ref{constraint}. Now the Hamiltonian becomes an
{\sl
  antiferromagnetic} transverse field Ising model on the dual triangular
lattice: \beqn H = \sum_{\bar{p}} - K \tau^x_{\bar{p}} +
\sum_{\bar{p},\bar{q}} J_{\bar{p},\bar{q}}
\tau^z_{\bar{p}}\tau^z_{\bar{q}} \eeqn For nearest neighbor sites
$\bar{p},\bar{q}$, $J_{\bar{p},\bar{q}} = h$. When
$J_{\bar{p},\bar{q}}$ dominates $K$, $\tau_{\bar{p}}^z$ takes on a
non-zero expectation value forming some pattern which optimizes
the $J_{\bar{p},\bar{q}}$ term. The non-zero ``condensate'' of
$\tau^z$ signals that the $Z_2$ gauge theory has entered the
confined phase.

The pattern of order in $\tau_{\bar{p}}^z$ depends upon the
detailed form of $J_{\bar{p},\bar{q}}$.  This can be analyzed by
treating $\tau^z_{\bar{p}}$ as a ``soft'' scalar field taking all
possible real values, rather than the integers $\pm 1$; this
approximation describes well the critical region in which
fluctuations on short time scales render the average of $\tau^z$
non-integral.  Then, the quadratic form defined by
$J_{\bar{p},\bar{q}}$ can be diagonalized in wavevector space and
generically has multiple minima in its Brillouin zone.  Physically
the eigenvalues of this quadratic form define the dispersion
relation of visons in the $Z_2$ phase.  On entering the confined
phase, the location of these minima determines the VBS pattern.
Notice that the physical VBS order parameter should always be a
bilinear of $\langle \tau^z \rangle $, since under transformation
$\tau^z \rightarrow - \tau^z$ the physical quantity $\sigma^x$ is
unchanged. In the following we will discuss four types of VBS
patterns on the honeycomb lattice.

\subsection{$c-$VBS order}

Now let us take the simplest case, with nonzero
$J_{\bar{p},\bar{q}}$ only between nearest neighbor dual sites
$\bar{p},\bar{q}$. Taking $h > 0$, the model becomes the nearest
neighbor frustrated quantum Ising model with transverse field.
This model was studied in Ref.~\cite{sondhi2001a}. Solving the
band structure of $\tau^z$, we find two inequivalent minima at the
corners of the vison BZ: $\vec{Q} = (\pm \frac{4\pi}{3}, 0)$.
Expanding $\tau^z$ at these two minima, we obtain a complex local
order parameter $\psi$: \beqn \tau^z \sim \psi
e^{i\frac{4\pi}{3}x} + \psi^\ast e^{- i\frac{4\pi}{3}x}. \eeqn The
low energy physics of visons should be fully characterized by
$\psi$.

Under discrete lattice symmetry, $\psi$ transforms as \beqn T_1
&:& x \rightarrow x + 1, \ \ \psi \rightarrow
e^{i\frac{4\pi}{3}}\psi, \cr\cr T_2 &:& x\rightarrow x +
\frac{1}{2}, \ y \rightarrow y + \frac{\sqrt{3}}{2}, \ \ \psi
\rightarrow e^{i\frac{2\pi}{3}}\psi, \cr\cr \mathrm{P}_y &:& x
\rightarrow -x, \ \ \psi \rightarrow \psi^\ast, \cr\cr
\mathrm{P}_x &:& y \rightarrow - y, \ \ \psi\rightarrow \psi,
\cr\cr \mathrm{T}  &:& t \rightarrow -t, \psi \rightarrow
\psi^\ast, \cr\cr \mathrm{R}_{\frac{2\pi}{3}} &:& \psi \rightarrow
\psi. \label{transformc}\eeqn $\mathrm{R}_{\frac{2\pi}{3}}$ is the
rotation by $2\pi/3$ around the center of hexagon.

The transformations in Eq.~\ref{transformc} determine that the low
energy Lagrangian for $\psi$ reads \beqn L = |\partial_\mu \psi|^2
+ r |\psi|^2 + u|\psi|^4 + w (\psi^6 + \psi^{\ast 6}), \label{XY}
\eeqn $i.e.$ The condensation of $\psi$ is described by a $3d$ XY
transition with $Z_6$ anisotropy. The physical VBS order parameter
$V$ should be a bilinear of $\psi$, $i.e.$ $V \sim \psi^2$. It is
straightforward to check that $V$ transforms in the same way as
the columnar VBS order parameter on the honeycomb lattice. Notice
that on the honeycomb lattice the $c-$VBS and the $\sqrt{3}\times
\sqrt{3}$ plaquette order have the same symmetry, hence the
condensate of $\psi$ can be either the $c-$VBS or the plaquette
order depending on the sign of $w$. Recently this plaquette order
has been observed with exact diagonalization on frustrated spin
models on the honeycomb lattice \cite{plaquette}. The $Z_6$
anisotropy introduced by the $w$ term in Eq.~\ref{XY} is an
irrelevant perturbation at the 3d XY fixed point.

%{\bf should
%one comment on the other possible pattern for the opposite sign of
%$w$ in the honeycomb case?}

If we approach this transition from the $c-$VBS side of the phase
diagram, this transition can be interpreted as a proliferation of
the vortex of $\psi$ $i.e.$ double vortex of VBS order paramter
$V$, while the single vortex of $V$ is still gapped. In fact, the
single vortex core of the $c-$VBS is attached with a spinon
(analogous to the square lattice case discussed in
Ref.~\cite{levinsenthil}), proliferation of single vortex will
lead to a spinon condensate, which corresponds to certain spin
order. However, if the spinon gap is finite, the finite
temperature thermal fluctuation can proliferate the single vortex.
Therefore although the quantum phase transition is driven by
double vortices, the finite temperature phase transition is still
driven by single vortex, hence at finite temperature the $Z_6$
anisotropy of Eq.~\ref{XY} becomes the $Z_3$ anisotropy, and there
is no algebraic Kosterlitz-Thouless phase at finite temperature.
This is a key difference between our current case and a physical
transverse field frustrated quantum Ising model, where a finite
temperature algebraic phase is expected \cite{sondhi2001a}.
%{\bf
%is
%  there some peculiarity reflecting the crossover from double to
%  single vortices at low but non-zero $T$?  What is the measureable
%  consequence here?}

\subsection{$s-$VBS order and four-fold plaquette order}

Now we modify the $Z_2$ gauge theory in Eq.~\ref{z2gauge} by
turning on the interaction between $Z_2$ electric field $\sigma^x$
on second nearest neighbor links: \beqn H_J = \sum_{\mathrm{2nd \
neighbor \ links}} J \sigma^x_{ij} \sigma^x_{kl}. \eeqn In the
dual theory this electric field interaction becomes a next nearest
neighbor hopping of $\tau^z$, and the full dual Hamiltonian reads
\beqn H = \sum_{\bar{p}} - K \tau^x_{\bar{p}} + \sum_{<
\bar{p},\bar{q} > } h \tau^z_{\bar{p}}\tau^z_{\bar{q}} + \sum_{\ll
\bar{p},\bar{q} \gg } J \tau^z_{\bar{p}}\tau^z_{\bar{q}}.
\label{jh} \eeqn The vison minima $(\pm \frac{4\pi}{3}, 0)$ are
stable with $J / h < 1/8$. When $1/8 < J / h < 1$, the minima of
the vison band structure are shifted to three inequivalent points
on the edges of BZ (Fig.~\ref{honeyvison1}$d$): \beqn && \vec{Q}_1
= (0, \frac{2\sqrt{3}\pi}{3}), \cr \cr && \vec{Q}_2 = ( - \pi, -
\frac{\sqrt{3}\pi}{3}), \cr \cr && \vec{Q}_3 = ( \pi, -
\frac{\sqrt{3}\pi}{3}). \label{honeysvbsq}\eeqn Notice that $-
\vec{Q}_a$ are equivalent to $\vec{Q}_a$ in the BZ.

Now three low energy modes can be defined by expanding $\tau^z$ at
momenta $\vec{Q}_a$: \beqn \tau^z \sim \sum_a \varphi_a \
e^{i\vec{Q}_a \cdot \vec{r}}. \eeqn Since $\vec{Q}_a$ and $-
\vec{Q}_a$ are equivalent, all three fields $\varphi_a$ are real.
Under lattice symmetry, $\varphi_a$ transform as \beqn T_1 &:&
\varphi_1 \rightarrow \varphi_1, \ \ \varphi_{2},\varphi_{3}
\rightarrow - \varphi_{2}, -\varphi_{3}, \cr\cr T_2 &:&
\varphi_{1},\varphi_{2} \rightarrow - \varphi_{1},-\varphi_{2}, \
\ \varphi_3 \rightarrow \varphi_3, \cr\cr \mathrm{P}_y &:&
\varphi_1 \rightarrow \varphi_1, \ \ \varphi_2, \varphi_3
\rightarrow \varphi_3, \varphi_2, \cr\cr \mathrm{P}_x &:&
\varphi_1 \rightarrow \varphi_1, \ \ \varphi_2, \varphi_3
\rightarrow \varphi_3, \varphi_2, \cr\cr \mathrm{T} &:& \varphi_a
\rightarrow \varphi_a, \cr\cr \mathrm{R}_{\frac{2\pi}{3}} &:&
\varphi_1 \rightarrow \varphi_2, \ \varphi_2 \rightarrow
\varphi_3, \ \varphi_3 \rightarrow \varphi_1.
\label{transforms}\eeqn

Now the symmetry allowed Lagrangian for $\varphi_a$ up to the
quartic order reads \beqn L = \sum_{a} (\partial_\mu \varphi_a)^2
+ r \varphi_a^2 + u (\sum_a \varphi_a^2)^2 + w (\sum_a \varphi_a^4
). \label{cubic}\eeqn This is an O(3) model with cubic anisotropy.
There are two possible types of condensates of $\varphi_a$:

({\it i}) When $w > 0$, the condensate $\langle \vec{\varphi}
\rangle$ are along the diagonal directions, and there are in total
four independent states with $\langle \vec{\varphi} \rangle \sim
(1, 1, 1)$, $(-1, -1, 1)$, $(-1, 1, -1)$, $(1, -1, -1)$. According
to the transformation of $\vec{\varphi}$, these four states
correspond to the four-sublattice plaquette phase
(Fig.~\ref{honeyvison1}$c$).

({\it ii}) When $w < 0$, the condensate
$\langle\vec{\varphi}\rangle$ has three fold degeneracy: $\langle
\vec{\varphi} \rangle \sim (1, 0, 0)$, $(0, 1, 0)$ and $(0, 0,
1)$. These three condensates break the rotation symmetry of the
lattice, but they do not break the translation symmetry. This is
again because physical order parameters are bilinears of
$\varphi_a$, hence they are insensitive to the sign change of
$\varphi_a$ under translation. These three states correspond
precisely to the three $s-$VBS pattern. Unlike the $c-$VBS, the
$s-$VBS is no longer described by an XY order parameter, and the
phase transition is not driven by vortex-like VBS defect.

The universality class of Eq.~\ref{cubic} was studied extensively
with $\epsilon = 4 - d$ expansion \cite{vicari2003}. The results
is that the O(3) Heisenberg fixed point is not stable. For $w > 0$
the transition is controlled by a stable cubic fixed point with
nonzero fixed point values $w^\ast$ and $u^\ast$.  This case
corresponds to the transition between the $Z_2$ spin liquid and
the four-sublattice plaquette phase described above.  For $w < 0$,
which corresponds to the transition to the $s-$VBS phase, there is
instead a run-away flow, which most likely implies a first order
transition. But if $w$ is small enough, for numerical simulations
on finite system the transition between $Z_2$ spin liquid and
$s-$VBS will be similar to the $3d$ O(3) transition.

Although we chose a specific vison hopping model Eq.~\ref{jh} to
obtain the vison band structure, the three minima $\vec{Q}_a$ in
the BZ are stable against any symmetry allowed perturbations on
Eq.~\ref{jh}. This is because no linear spatial derivative terms
are allowed in Eq.~\ref{cubic} by transformations
Eq.~\ref{transforms}.   Thus there are only two ways to
destabilize the minima in the BZ: ({\it 1}). the current minima
will be replaced by a new set of minima through a first order
transition, like the transition between $c-$VBS and $s-$VBS at
$J/h = 1/8$; ({\it 2}). the sign of the spatial derivative terms
in Eq.~\ref{jh} changes through a second order Lifshitz
transition. The second situation will lead to the incommensurate
VBS order, which will be discussed in the next subsection.

\subsection{incommensurate VBS}

In Eq.~\ref{jh}, if $J / h > 1$, the vison band structure has six
inequivalent incommensurate minima $\vec{Q}_a$ in its BZ
(Fig.~\ref{honeyvison1}$d$). Since $\vec{Q}_a$ and $- \vec{Q}_a$
are no longer equivalent, the low energy vison modes are described
by three {\sl complex} fields $\varphi_a$: \beqn \tau^z \sim
\sum_a \left(
  \varphi_a e^{i \vec{Q}_a \cdot \vec{r}} + \varphi_a^\ast e^{ - i
    \vec{Q}_a \cdot \vec{r}} \right). \eeqn Under lattice symmetries,
these vison fields transform as \beqn T_1 &:& \varphi_a
\rightarrow e^{i\vec{Q}_{a,x} x}, \cr\cr \mathrm{P}_y &:&
\varphi_1 \rightarrow \varphi_1, \ \ \varphi_2, \varphi_3
\rightarrow \varphi_3, \varphi_2, \cr\cr \mathrm{P}_x &:&
\varphi_1 \rightarrow \varphi_1^\ast, \ \ \varphi_2, \varphi_3
\rightarrow \varphi^\ast_3, \varphi^\ast_2, \cr\cr \mathrm{T} &:&
\varphi_a \rightarrow \varphi^\ast_a, \cr\cr
\mathrm{R}_{\frac{2\pi}{3}} &:& \varphi_1 \rightarrow \varphi_2, \
\varphi_2 \rightarrow \varphi_3, \ \varphi_3 \rightarrow
\varphi_1. \label{transformi}\eeqn The field theory describing
$\varphi_a$ is \beqn L &=& \sum_{a} |\partial_\mu \varphi_a|^2 + r
|\varphi_a|^2 + u (\sum_a |\varphi_a|^2)^2 + w (\sum_a
|\varphi_a|^4 ) \cr\cr & + & v_1 |\varphi_1|^2 |\varphi_2|^2
|\varphi_3|^2 + v_2 (\varphi_1^2 \varphi_2^2 \varphi_3^2 + H.c.) +
\cdots \eeqn When $\vec{\varphi}$ condenses, the system exhibits
incommensurate VBS order. Since incommensurate VBS order has yet
to be observed numerically, we will not explore this phase or the
corresponding transitions further.

\subsection{A quantum dimer model}

In addition to the numerical works introduced in section
\ref{introduction}, a recent QMC simulation discovered a gapped
liquid phase in the SU(4) Hubbard model on the honeycomb lattice.
In those simulations, this liquid phase appeared between the
semimetal phase and $c-$VBS phase, and SU(4) N\'{e}el order was
completely absent \cite{assaad2010c}. Since all the gapped liquids
observed numerically are adjacent to VBS phases, it is natural to
approach the gapped liquid phase starting with dimer (spin
singlet) basis.  In this subsection we write down a quantum dimer
model which realizes part of the physics discussed above.

\begin{figure}
\includegraphics[width=2.7 in]{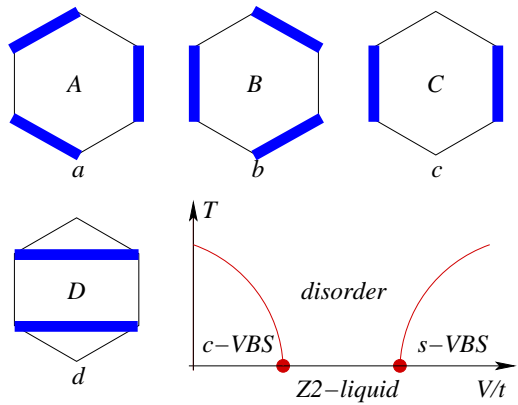}
\caption{({\it a}) $-$ ({\it d}), the dimer patterns involved in
the quantum dimer model Eq.~\ref{QDM0} and Eq.~\ref{QDM1}. {\it
e}, phase diagram of the quantum dimer model tuned with $V/t$,
with the presence of $H_1$ in Eq.~\ref{QDM1}. }
\label{honeyvison2}
\end{figure}

The standard quantum dimer model (QDM) on the honeycomb lattice
has been studied carefully in the past \cite{fradkin2004}: \beqn
&& H_0 = H_t + H_v \cr\cr &=& - t \ (|A\rangle \langle B| +
|B\rangle \langle A|) + V \ (|A\rangle \langle A| + |B\rangle
\langle B|). \label{QDM0}\eeqn $|A\rangle$ and $|B\rangle$ are
dimer configurations depicted in Fig.~\ref{honeyvison2}$a$, $b$.
This model can be mapped to a compact U(1) gauge theory in the
standard way: \beqn \vec{e}_{ij} = \vec{v} \cdot n_{ij}, \ \ H_t
\sim - t \cos(\vec{\nabla} \times \vec{a}). \eeqn Here $n_{ij} =
0, 1$ is the dimer density on each link $(i,j)$, $\vec{v}$ is a
unit vector defined on each link, and it always points from
sublattice A to B on the honeycomb lattice. $\vec{e}$ and
$\vec{a}$ are electric field and gauge vector potential
respectively. Due to the confinement of compact U(1) gauge theory
in two dimensions \cite{polyakovbook}, the system is gapped and
VBS ordered throughout the phase diagram, except for the isolated
gapless RK point at $V = t$.  There is thus no stable gapped
liquid phase in this phase diagram \cite{fradkin2004}. Therefore
in order to understand the gapped liquid phase around VBS phases
on the honeycomb lattice, a modified QDM is needed.

First of all, we notice that all the gapped liquid phases observed
numerically only occur at intermediate $J_2/J_1$ or $t/U$
(Ref.~\cite{claire2001,clark2010,assaad2010c}).  In this regime,
there must be a considerable probability for dimers to form not
only between nearest neighbor sites but also between next nearest
neighbor sites. Then, in addition to the standard QDM
Eq.~\ref{QDM0}, we turn on the following dimer flipping term:
\beqn H_1 = - \tilde{t} \ (|C\rangle \langle D| + |D\rangle
\langle C|). \label{QDM1}\eeqn $|C\rangle$ and $|D\rangle$ are
dimer configurations in Fig.~\ref{honeyvison2}$c$, $d$. This term
breaks the U(1) gauge symmetry of the original QDM down to $Z_2$
gauge symmetry, because it annihilates two electric flux quanta
along the same direction. An alternative way of understanding this
$Z_2$ effect is by noticing that annihilating configuration
$|C\rangle$ is equivalent to hopping two unit U(1) gauge charges,
hence $H_1$ is equivalent to the following term at low energy:
\beqn H_1 \sim - \cos(2\nabla_\mu \phi - 2 a_\mu).
\label{double}\eeqn Here $\exp(i\phi)$ creates a unit gauge
charge. When $\phi$ condenses, it breaks the U(1) gauge symmetry
down to $Z_2$. As an analogue, the triangular lattice QDM can also
be viewed as a square lattice QDM with extra diagonal dimers,
which also breaks the U(1) gauge symmetry down to $Z_2$.

The term $H_1$ can drive the system into a $Z_2$ liquid phase on
the honeycomb lattice.  In the presence of non-zero $H_1$, a
possible schematic phase diagram for the QDM is shown in
Fig.~\ref{honeyvison2}$e$.  The $Z_2$ liquid phase intervenes
between the $c-$VBS and $s-$VBS phases. The nature of the phase
transitions in this phase diagram have been discussed in
subsection IIA and IIB.

The transition between the $Z_2$ liquid phase and the $c-$VBS can
also be more intuitively understood as follows: The $c-$VBS is a
confined phase of the compact U(1) gauge field, where the
triple-monopole of the compact U(1) gauge field leads to the
following height field theory in the dual theory: \beqn L_h =
(\partial_\mu h)^2 - g \cos(6\pi h), \eeqn where height field $h$
is defined as $\vec{e} = \hat{z} \times \vec{\nabla} h$, and
$\exp(i 2\pi h)$ creates a flux quantum of the compact U(1) gauge
field $\vec{a}$. The vertex operator $- g\cos(6\pi h)$ has three
independent minima $h = 0, \ \pm 1/3$, which corresponds to the
three fold degenerate $c-$VBS phase. In the $Z_2$ liquid phase,
since the U(1) gauge symmetry is broken down to $Z_2$, there are
stable $\pi$-flux excitations of the U(1) gauge field. Hence the
triple-monopole of the U(1) gauge field corresponds to
creating/annihilating six $\pi$-flux. If we describe the
transition in terms of these $\pi$-flux excitations, the field
theory takes exactly the same form as Eq.~\ref{XY}.

At the transition between the $c-$VBS and $Z_2$ liquid phase,
based on the well-known critical exponent of the 3d XY fixed point
obtained from various methods \cite{vicari2003}, we predict the
anomalous dimension of the $c-$VBS order parameter to be $\eta_{V}
= 1.47$ (much larger than the ordinary Wilson-Fisher transitions).
Also, close to this transition, the $c-$VBS order parameter scales
as $\langle V \rangle \sim (r_c - r)^{\beta}$, with $\beta =
0.83$. These predictions can be checked numerically.

\section{Odd $Z_2$ gauge theory on the square lattice} \label{z2square}

\begin{figure}
\includegraphics[width=3.0 in]{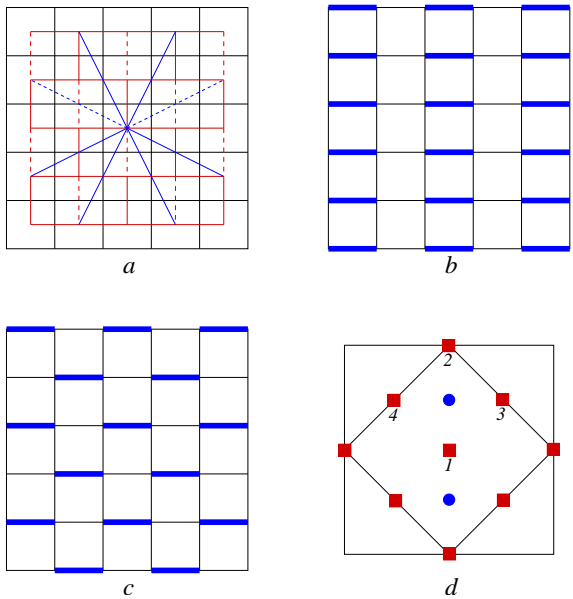}
\caption{({\it a}), the dual square lattice. The vison (vortex)
hopping on the dashed bonds are negative. ({\it b}), ({\it c}),
the $c-$VBS and $s-$VBS patterns. ({\it d}), the vison (vortex)
Brillouin zone. When the nearest neighbor vison (vortex) hopping
is dominant, there are two inequivalent minima located at $(0, \pm
\frac{\pi}{2})$ (circles); when the 4th neighbor hopping is
dominant, there are four inequivalent minima described by
Eq.~\ref{squaresvbsq}.} \label{squarevison1}
\end{figure}

Now we switch gears to the odd $Z_2$ gauge theory on the square
lattice. Again we want to discuss the phase transition from the
$Z_2$ liquid phase to both $c-$VBS and $s-$VBS. The odd $Z_2$
gauge theory is dual to a transverse field quantum Ising model on
the dual square lattice (Fig.~\ref{squarevison1}$a$).  Unlike the
honeycomb lattice case, now the dual quantum Ising model has to
apparently break the lattice symmetry in any specific gauge
choice.  The correct lattice symmetry transformation for the dual
vison field $\tau^z$ must be combined with a nontrivial $Z_2$
gauge transformation, $i.e.$ $\tau^z$ carries a projective
representation of the symmetry group. The dual quantum Ising model
has to be invariant under the projective symmetry group (PSG).

We consider the following Hamiltonian for the dual Ising model:
\beqn H = \sum_{\bar{p} }- K \tau^x_{\bar{p}} +
\sum_{<\bar{p},\bar{q}>}
J_{\bar{p},\bar{q}}\tau^z_{\bar{p}}\tau^z_{\bar{q}} +
\sum_{\bar{p},\bar{q}}
J^\prime_{\bar{p},\bar{q}}\tau^z_{\bar{p}}\tau^z_{\bar{q}}.
\label{squareising}\eeqn $J$ and $J^\prime$ denote the nearest and
fourth nearest neighbor Ising couplings. $J$ and $J^\prime$ are
chosen to be positive on all the solid bonds, but negative on all
the dashed bonds in Fig.~\ref{squarevison1}$a$. The Hamiltonian of
Eq.~\ref{squareising} with the current choice of gauge is
invariant under the PSG of $\tau^z$. Notice that 2nd nearest
neighbor Ising couplings are entirely prohibited by the PSG.

If $J^\prime / J < 0.0858$, there are two inequivalent minima in
the vison band structure, located at $\vec{Q} = (0, \pm
\frac{\pi}{2})$. Again we can expand $\tau^z$ at these two minima
as \beqn \tau^z \sim \varphi e^{i\frac{\pi}{2}y} + \varphi^\ast
e^{- i\frac{\pi}{2}y}. \eeqn The PSG for $\varphi$ reads \beqn
\mathrm{T}_x &:& x \rightarrow x+1, \ \ \varphi \rightarrow
e^{i\frac{\pi}{4} x}\varphi^\ast,  \cr\cr \mathrm{T}_y &:& y
\rightarrow y + 1, \ \ \varphi \rightarrow e^{- i\frac{\pi}{4}
x}\varphi^\ast, \cr\cr \mathrm{P}_y &:& x \rightarrow -x, \ \
\varphi \rightarrow \varphi, \cr\cr \mathrm{P}_x &:& y \rightarrow
-y, \ \ \varphi \rightarrow \varphi, \cr\cr \mathrm{P}_{x+y} &:& x
\rightarrow y, \ y \rightarrow x, \ \ \varphi \rightarrow
i\varphi^ \ast. \eeqn Notice that the reflection $\mathrm{P}_x$
and $\mathrm{P}_y$ are site-centered reflection of the dual
lattice (bond-centered reflection of the original lattice). The
PSG allowed field theory for $\varphi$ reads \beqn L =
|\partial_\mu \varphi|^2 + r |\varphi|^2 + g|\varphi|^4 + w
(\varphi^8 + \varphi^{\ast 8}). \label{z8} \eeqn The gauge
invariant physical order parameters are \beqn c-\mathrm{VBS}_x &:&
e^{i\frac{\pi}{4}}\varphi^2 + e^{- i\frac{\pi}{4}}\varphi^{\ast
2}, \cr\cr c-\mathrm{VBS}_y &:& e^{- i\frac{\pi}{4}}\varphi^2 +
e^{i\frac{\pi}{4}}\varphi^{\ast 2}. \eeqn The quantum phase
transition between the $Z_2$ liquid and the $c-$VBS is a $3d$ XY
transition, since the $Z_8$ anisotropy in Eq.~\ref{z8} is highly
irrelevant at the $3d$ XY fixed point. This result is consistent
with previous studies on fully frustrated Ising model on the cubic
lattice \cite{ffisingcubic1,ffisingcubic2}.

When $J^\prime / J > 0.0858$, the minima of the vison band
structure are shifted to four other inequivalent momenta in the
BZ: \beqn && Q_1 = (0,0), \ \ Q_2 = (0, \pi), \cr\cr && Q_3 =
(\frac{\pi}{2}, \frac{\pi}{2}), \ \ Q_4 = (- \frac{\pi}{2},
\frac{\pi}{2}). \label{squaresvbsq} \eeqn We denote the low energy
vison modes at these four momenta by $\varphi_a$ with $a = 1
\cdots 4$. Notice all these four modes are real fields, because
$Q_a$ are equivalent to $- Q_a$.

The PSG action on $\varphi_a$ reads \beqn \mathrm{T}_x &:&
\varphi_1 \rightarrow \varphi_2, \ \ \varphi_2 \rightarrow
\varphi_1, \ \ \varphi_3 \rightarrow \varphi_4, \ \ \varphi_4
\rightarrow - \varphi_3, \cr\cr \mathrm{T}_y &:& \varphi_1
\rightarrow \varphi_2, \ \ \varphi_2 \rightarrow -\varphi_1, \ \
\varphi_3 \rightarrow \varphi_4, \ \ \varphi_4 \rightarrow
\varphi_3, \cr\cr \mathrm{P}_y &:& \varphi_1 \rightarrow
\varphi_1, \ \ \varphi_2 \rightarrow \varphi_2, \ \ \varphi_3
\rightarrow \varphi_4, \ \ \varphi_4 \rightarrow \varphi_3, \cr\cr
\mathrm{P}_x &:& \varphi_1 \rightarrow - \varphi_2, \ \ \varphi_2
\rightarrow - \varphi_1, \ \ \varphi_3 \rightarrow \varphi_3, \ \
\varphi_4 \rightarrow \varphi_4,  \cr\cr \mathrm{P}_{x+y} &:&
\varphi_1 \rightarrow \varphi_3, \ \ \varphi_3 \rightarrow
\varphi_1, \ \ \varphi_2 \rightarrow \varphi_4, \ \ \varphi_4
\rightarrow \varphi_2. \eeqn

It is straightforward to show that these four vison minima
actually describe the $s-$VBS pattern on the square lattice. By
using the PSG transformations above, we find that fields
transforming as the $s-$VBS order parameters are \beqn
s-\mathrm{VBS}_x &:& \varphi_1^2 - \varphi_2^2, \cr\cr
s-\mathrm{VBS}_y &:& \varphi_3^2 - \varphi_4^2. \eeqn

Applying the PSG to obtain a general invariant field theory, we
find, up to quartic order the Lagrangian \beqn L &=& \sum_{a}
(\partial_\mu \varphi_a)^2 + r \varphi_a^2 + u (\sum_a
\varphi_a^2)^2 \cr\cr &+& w (\sum_a \varphi_a^4 ) + v (\varphi_1^2
+ \varphi_2^2)(\varphi_3^2 + \varphi_4^2) \label{o4} \eeqn The
location of vison band structure minima is stable, since no linear
spatial derivative terms are allowed in this field theory. The
first line of this equation describes an O(4) theory, which is
very different from the effective XY theory in the $c-$VBS case.
The second line of Eq.~\ref{o4} breaks this O(4) symmetry down to
$Z_4 \times Z_4 \times Z_2$. When $v > 0$, $w < 0$, the visons
condense in a way that yields $s-$VBS order.  However, according
to the high order $\epsilon$ expansion in Ref.~\cite{vicari2003b},
both $v$ and $w$ are relevant perturbations at the 3d O(4)
universality class.  This likely indicates the lack of a direct
continuous $Z_2$ spin liquid to s-VBS transition (though a weakly
first-order transition would be possible).

\section{Magnetic order $-$ VBS transitions} \label{ordervbs}

The phase transition between the standard collinear ordered
antiferromagnet and the $c-$VBS phase is described by the theory
of deconfined criticality, with the critical effective field
theory being the noncompact CP(1) model
\cite{senthil2004,senthil2004a}. This deconfined phase transition
is realized in the $J-Q$ model on the square lattice with both two
spin and four spin interactions \cite{sandvik2007}\cite{kaul2007}.
Recent exact diagonalization simulation of the same model (with
full parameter space of $J-Q$) \cite{claire2004} and QMC on a
modified $J-Q$ model \cite{sandvik2009} discovered that there can
also be a direct transition between the N\'{e}el and $s-$VBS
order, and this transition is what we will try to understand in
this section.

To simplify this problem, we turn on an easy plane anisotropy on
the spin system. Now the spin-1/2 problem is equivalent to a
hard-core boson model at half-filling, and the N\'{e}el order is
mapped to the superfluid phase of the boson system. It is
well-known that the hard-core boson problem is dual to its vortex
theory, where vortices are bosons hopping on the dual lattice
sites, with coupling to a dynamical U(1) gauge field, this U(1)
gauge field is precisely the dual of the Goldstone mode of the
superfluid phase. Because the boson is half-filled, a vortex will
see a $\pi-$flux through each of the dual plaquettes.

Previous studies showed that if the nearest neighbor vortex
hopping is considered in the dual Hamiltonian, the transition
between superfluid (N\'{e}el order) and $c-$VBS may be described
as the condensation of vortices
\cite{balentsvortex,balentsburkov}. In the follows we will see
that if further neighbor vortex hoppings are taken into account,
the superfluid (N\'{e}el order) to $s-$VBS transition can also be
understood as vortices condensing in its BZ, just like the vison
theory in the previous sections.

For instance, on both the honeycomb and square lattice, in order
to describe the $s-$VBS phase, we only need to turn on  exactly
the same further neighbor vortex hopping as the previous vison
theories in section \ref{z2honeycomb} and \ref{z2square}. The PSG
of the vortices is almost identical to that of the visons in the
previous sections, thus we will not write it down explicitly. The
main differences between the vortex theory and vison theory are,
({\it 1}) the vortex is described by a complex boson instead of
real boson, ({\it 2}) the vortex is coupled to a dynamical U(1)
gauge field, ({\it 3}) physical order parameters should be U(1)
gauge invariant.

We summarize our results here:

({\it i}) on the honeycomb lattice, with 2nd neighbor vortex
hopping, the vortex band structure can have minima located at the
three momenta in Eq.~\ref{honeysvbsq}, the low energy vortex mode
at each momentum is described by a complex boson field
$\varphi_a$. The transition between the superfluid (magnetic
order) and the $s-$VBS is described by the CP(2) theory with cubic
anisotropy: \beqn L & = & \sum_{\alpha = 1}^3 |(\partial_\mu -
iA_\mu )\varphi_\alpha|^2 + r|\varphi_\alpha|^2 + g(\sum_\alpha
|\varphi_\alpha|^2)^2 \cr\cr & + & \sum_\alpha u
|\varphi_\alpha|^4 + \cdots \label{honeyneelsvbs}\eeqn When $u <
0$ and $r<0$, only one of the vortex modes is condensed, which
corresponds to the $s-$VBS order. Although $\varphi_a$ is a
complex field, its condensate has no gapless Goldstone mode, due
to its coupling to U(1) gauge field $A_\mu$.  It is unclear
whether cubic anisotropy is relevant or not at this CP(2) quantum
critical point. The scaling dimension of the cubic anistropy can
be calculated systematically with the standard $1/N$ expansion.
%but this issue is important to understanding the
%stability of this critical theory.

({\it ii}) on the square lattice, the transition between the
superfluid (magnetic order) and the $s-$VBS is described by the
CP(3) theory with anisotropies that break the SU(4) symmetry down
to $(\mathrm{U(1)})^4$ and discrete interchange symmetries between
$\varphi_a$: \beqn L & = & \sum_{\alpha = 1}^4 |(\partial_\mu -
iA_\mu )\varphi_\alpha|^2 + r|\varphi_\alpha|^2 + g(\sum_\alpha
|\varphi_\alpha|^2)^2 \cr\cr & + & \sum_\alpha u
|\varphi_\alpha|^4 + v (|\varphi_1|^2 +
|\varphi_2|^2)(|\varphi_3|^2 + |\varphi_4|^2) + \cdots
\label{squareneelsvbs}\eeqn $\varphi_\alpha$ are four low energy
vortex modes at the four momenta in Eq.~\ref{squaresvbsq}. When $u
< 0$, $v > 0$ the condensate of $\varphi_a$ corresponds to the
$s-$VBS order. The $s-$VBS order parameters are \beqn
s-\mathrm{VBS}_x &:& |\varphi_1|^2 - |\varphi_2|^2, \cr\cr
s-\mathrm{VBS}_y &:& |\varphi_3|^2 - |\varphi_4|^2.
\label{svbsop}\eeqn Here there are two quartic anisotropies,
though again their scaling behavior at the CP(3) point needs to be
addressed by further calculations.

It was shown in Ref.~\cite{levinsenthil} that the $Z_4$ vortex of
$c-$VBS has to carry a spinon. However, in Fig.~\ref{sVBSdefect}
we illustrated that there is no spinon attached to the $Z_4$
vortex of the $s-$VBS, thus the $s-$VBS to magnetic order
transition should not be induced with $Z_4$ vortex proliferation.
This difference can be understood with the $s-$VBS order
parameters in Eq.~\ref{svbsop}: the flux of the U(1) gauge field
in %Eq.~\ref{honeyneelsvbs} and
Eq.~\ref{squareneelsvbs} carries spin (since $A_\mu$ is the dual
of Goldstone mode associated with $S^z$ conservation), while the
$Z_4$ vortex of $s-$VBS is simply a vortex surrounded by
$\varphi_1$, $\varphi_3$, $\varphi_2$ and $\varphi_4$ condensate
cyclicly, thus no gauge flux is attached with this vortex core. In
fact, the U(1) gauge flux corresponds to the ``skyrmion" type of
defect of the CP(2) and CP(3) manifold. Thus to induce a
magnetically ordered phase, we need to condense the more
complicated skyrmion defect of the $s-$VBS order.

\section{Summary and Outlook}
\label{sec:summary-outlook}

In this work we studied the quantum transitions from the $Z_2$
spin liquid and magnetically ordered phases to different types of
VBS orders, with the focus on the staggered VBS. Although the
$s-$VBS and $c-$VBS have the same degeneracy, the low energy field
theories describing these two cases close to the transitions are
very different. In our current work we only considered the case of
transitions from magnetically ordered phases with easy plane
anisotropy, while a description for the SU(2) invariant case is
still needed. We will leave this case to future study.  Moreover,
even in the easy plane case, the field theories we obtained are
not well-studied, and the effects of anisotropies on their
critical properties (and indeed stability of the latter) deserve
more detailed study.

%In Ref.~\cite{claire2001,clark2010}, it was shown that the liquid
%phase of the $J_1 - J_2$ spin model on the honeycomb lattice is
%sandwiched between the $s-$VBS and the N\'{e}el order. If we
%assume the observed liquid phase is indeed the $Z_2$ liquid phase,
%then the transition between the liquid and $s-$VBS was discussed
%in this paper. However, the transition between the $Z_2$ liquid
%and N\'{e}el order is even more nontrivial. It is tempting to
%interpret this transition as condensing the SU(2) spinon or
%Schwinger boson $z_\alpha$ in the $Z_2$ liquid phase. However,
%because $z_\alpha$ is coupled with a $Z_2$ gauge field in the
%liquid phase, the spinon condensate will have ground state
%manifold SO(3) instead of $S^2$, thus the spinon condensate is
%different from the N\'{e}el order. In order to reconcile this
%inconsistency, Ref.~\cite{ran2010b,ran2010} and Ref.~\cite{xu2010}
%proposed different types of hidden orders in the N\'{e}el order
%close to the liquid phase, which await numerical verifications.

The formalisms we used in this work, namely the odd $Z_2$ gauge
theory, the dual vortex theory, and quantum dimer model, are all
theories for strongly coupled systems, $i.e.$ the electric charge
excitations of this system were ignored completely. However, the
deconfined quantum criticality theory can also be formulated in
the weak coupling limit, as it can be interpreted as the phase
transition between five competing mass gap order parameters of
Dirac fermion \cite{senthil2007}. The Dirac fermion will generate
a topological Wess-Zumino-Witten (WZW) term for these competing
orders \cite{abanov2000,abanov2001}, and physically the WZW term
attaches a spin-1/2 degree of freedom to each vortex core of the
$c-$VBS. Fermions on both honeycomb lattice, and square lattice
with $\pi-$flux through every plaquette have Dirac fermion in the
band structure, and in both cases the N\'{e}el and $c-$VBS are
five competing order parameters with a WZW term \cite{ryuwzw}.

Unlike the $c-$VBS, the $s-$VBS does not generate a Dirac mass gap
for either honeycomb or square lattice $\pi-$flux state, thus the
standard calculation used in Ref.~\cite{abanov2000,abanov2001}
does not lead to a WZW term between $s-$VBS and N\'{e}el order.
This observation also suggests that the $s-$VBS and $c-$VBS are
fundamentally different, which echoes the results obtained in our
current paper. In the future, it would be very meaningful to also
pursue a weak coupling version of the theory for quantum phase
transitions with $s-$VBS using fermion band structure.

%\begin{acknowledgments}
L.B. was supported by NSF grants DMR-0804564 and PHY05-51164.
%\end{acknowledgments}

\bibliography{honeyvison}

\begin{thebibliography}{31}
\expandafter\ifx\csname natexlab\endcsname\relax\def\natexlab#1{#1}\fi
\expandafter\ifx\csname bibnamefont\endcsname\relax
  \def\bibnamefont#1{#1}\fi
\expandafter\ifx\csname bibfnamefont\endcsname\relax
  \def\bibfnamefont#1{#1}\fi
\expandafter\ifx\csname citenamefont\endcsname\relax
  \def\citenamefont#1{#1}\fi
\expandafter\ifx\csname url\endcsname\relax
  \def\url#1{\texttt{#1}}\fi
\expandafter\ifx\csname urlprefix\endcsname\relax\def\urlprefix{URL }\fi
\providecommand{\bibinfo}[2]{#2}
\providecommand{\eprint}[2][]{\url{#2}}

\bibitem[{\citenamefont{Senthil
  et~al.}(2004{\natexlab{a}})\citenamefont{Senthil, Balents, Sachdev,
  Vishwanath, and Fisher}}]{senthil2004a}
\bibinfo{author}{\bibfnamefont{T.}~\bibnamefont{Senthil}},
  \bibinfo{author}{\bibfnamefont{L.}~\bibnamefont{Balents}},
  \bibinfo{author}{\bibfnamefont{S.}~\bibnamefont{Sachdev}},
  \bibinfo{author}{\bibfnamefont{A.}~\bibnamefont{Vishwanath}},
  \bibnamefont{and} \bibinfo{author}{\bibfnamefont{M.~P.~A.}
  \bibnamefont{Fisher}}, \bibinfo{journal}{Phys. Rev. B}
  \textbf{\bibinfo{volume}{70}}, \bibinfo{pages}{144407}
  (\bibinfo{year}{2004}{\natexlab{a}}).

\bibitem[{\citenamefont{Senthil
  et~al.}(2004{\natexlab{b}})\citenamefont{Senthil, Vishwanath, Balents,
  Sachdev, and Fisher}}]{senthil2004}
\bibinfo{author}{\bibfnamefont{T.}~\bibnamefont{Senthil}},
  \bibinfo{author}{\bibfnamefont{A.}~\bibnamefont{Vishwanath}},
  \bibinfo{author}{\bibfnamefont{L.}~\bibnamefont{Balents}},
  \bibinfo{author}{\bibfnamefont{S.}~\bibnamefont{Sachdev}}, \bibnamefont{and}
  \bibinfo{author}{\bibfnamefont{M.~P.~A.} \bibnamefont{Fisher}},
  \bibinfo{journal}{Science} \textbf{\bibinfo{volume}{303}},
  \bibinfo{pages}{1409} (\bibinfo{year}{2004}{\natexlab{b}}).

\bibitem[{\citenamefont{Haldane}(1988)}]{haldanemonopole}
\bibinfo{author}{\bibfnamefont{F.~D.~M.} \bibnamefont{Haldane}},
  \bibinfo{journal}{Physical Review Letter} \textbf{\bibinfo{volume}{61}},
  \bibinfo{pages}{1029} (\bibinfo{year}{1988}).

\bibitem[{\citenamefont{Read and Sachdev}(1990)}]{sachdev1990}
\bibinfo{author}{\bibfnamefont{N.}~\bibnamefont{Read}} \bibnamefont{and}
  \bibinfo{author}{\bibfnamefont{S.}~\bibnamefont{Sachdev}},
  \bibinfo{journal}{Phys. Rev. B} \textbf{\bibinfo{volume}{42}},
  \bibinfo{pages}{4568} (\bibinfo{year}{1990}).

\bibitem[{\citenamefont{Lou et~al.}(2009)\citenamefont{Lou, Sandvik, and
  Kawashima}}]{sandvik2009}
\bibinfo{author}{\bibfnamefont{J.}~\bibnamefont{Lou}},
  \bibinfo{author}{\bibfnamefont{A.~W.} \bibnamefont{Sandvik}},
  \bibnamefont{and}
  \bibinfo{author}{\bibfnamefont{N.}~\bibnamefont{Kawashima}},
  \bibinfo{journal}{Phys. Rev. B} \textbf{\bibinfo{volume}{80}},
  \bibinfo{pages}{180414(R)} (\bibinfo{year}{2009}).

\bibitem[{\citenamefont{Mulder et~al.}(2010)\citenamefont{Mulder, Ganesh,
  Capriotti, and Paramekanti}}]{paramekantisvbs}
\bibinfo{author}{\bibfnamefont{A.}~\bibnamefont{Mulder}},
  \bibinfo{author}{\bibfnamefont{R.}~\bibnamefont{Ganesh}},
  \bibinfo{author}{\bibfnamefont{L.}~\bibnamefont{Capriotti}},
  \bibnamefont{and}
  \bibinfo{author}{\bibfnamefont{A.}~\bibnamefont{Paramekanti}},
  \bibinfo{journal}{Phys. Rev. B} \textbf{\bibinfo{volume}{81}},
  \bibinfo{pages}{214419} (\bibinfo{year}{2010}).

\bibitem[{\citenamefont{Fouet et~al.}(2001)\citenamefont{Fouet, Sindzingre, and
  Lhuillier}}]{claire2001}
\bibinfo{author}{\bibfnamefont{J.~B.} \bibnamefont{Fouet}},
  \bibinfo{author}{\bibfnamefont{P.}~\bibnamefont{Sindzingre}},
  \bibnamefont{and}
  \bibinfo{author}{\bibfnamefont{C.}~\bibnamefont{Lhuillier}},
  \bibinfo{journal}{Euro. Phys. Journal. B} \textbf{\bibinfo{volume}{20}},
  \bibinfo{pages}{241} (\bibinfo{year}{2001}).

\bibitem[{\citenamefont{Clark et~al.}(2010)\citenamefont{Clark, Abanin, and
  Sondhi}}]{clark2010}
\bibinfo{author}{\bibfnamefont{B.~K.} \bibnamefont{Clark}},
  \bibinfo{author}{\bibfnamefont{D.~A.} \bibnamefont{Abanin}},
  \bibnamefont{and} \bibinfo{author}{\bibfnamefont{S.~L.} \bibnamefont{Sondhi}}
  (\bibinfo{year}{2010}), \eprint{arXiv:1010.3011}.

\bibitem[{\citenamefont{Reuther et~al.}(2011)\citenamefont{Reuther, Abanin, and
  Thomale}}]{thomalehoney}
\bibinfo{author}{\bibfnamefont{J.}~\bibnamefont{Reuther}},
  \bibinfo{author}{\bibfnamefont{D.}~\bibnamefont{Abanin}}, \bibnamefont{and}
  \bibinfo{author}{\bibfnamefont{R.}~\bibnamefont{Thomale}}
  (\bibinfo{year}{2011}), \eprint{arXiv:1103.0859}.

\bibitem[{\citenamefont{Laeuchli et~al.}(2004)\citenamefont{Laeuchli, Domenge,
  Lhuillier, Sindzingre, and Troyer}}]{claire2004}
\bibinfo{author}{\bibfnamefont{A.}~\bibnamefont{Laeuchli}},
  \bibinfo{author}{\bibfnamefont{J.~C.} \bibnamefont{Domenge}},
  \bibinfo{author}{\bibfnamefont{C.}~\bibnamefont{Lhuillier}},
  \bibinfo{author}{\bibfnamefont{P.}~\bibnamefont{Sindzingre}},
  \bibnamefont{and} \bibinfo{author}{\bibfnamefont{M.}~\bibnamefont{Troyer}},
  \bibinfo{journal}{Phys. Rev. Lett.} \textbf{\bibinfo{volume}{95}},
  \bibinfo{pages}{137206} (\bibinfo{year}{2004}).

\bibitem[{\citenamefont{Sen and Sandvik}(2010)}]{sandvik2009a}
\bibinfo{author}{\bibfnamefont{J.}~\bibnamefont{Sen}} \bibnamefont{and}
  \bibinfo{author}{\bibfnamefont{A.~W.} \bibnamefont{Sandvik}},
  \bibinfo{journal}{Phys. Rev. B} \textbf{\bibinfo{volume}{82}},
  \bibinfo{pages}{174428} (\bibinfo{year}{2010}).

\bibitem[{\citenamefont{Meng et~al.}(2010)\citenamefont{Meng, Lang, Wessel,
  Assaad, and Muramatsu}}]{meng}
\bibinfo{author}{\bibfnamefont{Z.~Y.} \bibnamefont{Meng}},
  \bibinfo{author}{\bibfnamefont{T.~C.} \bibnamefont{Lang}},
  \bibinfo{author}{\bibfnamefont{S.}~\bibnamefont{Wessel}},
  \bibinfo{author}{\bibfnamefont{F.~F.} \bibnamefont{Assaad}},
  \bibnamefont{and}
  \bibinfo{author}{\bibfnamefont{A.}~\bibnamefont{Muramatsu}},
  \bibinfo{journal}{Nature} \textbf{\bibinfo{volume}{464}},
  \bibinfo{pages}{847} (\bibinfo{year}{2010}).

\bibitem[{\citenamefont{Wen}(2002)}]{wen2002a}
\bibinfo{author}{\bibfnamefont{X.~G.} \bibnamefont{Wen}},
  \bibinfo{journal}{Phys. Rev. B} \textbf{\bibinfo{volume}{65}},
  \bibinfo{pages}{165113} (\bibinfo{year}{2002}).

\bibitem[{\citenamefont{Moessner and Sondhi}(2001)}]{sondhi2001a}
\bibinfo{author}{\bibfnamefont{R.}~\bibnamefont{Moessner}} \bibnamefont{and}
  \bibinfo{author}{\bibfnamefont{S.~L.} \bibnamefont{Sondhi}},
  \bibinfo{journal}{Phys. Rev. B} \textbf{\bibinfo{volume}{63}},
  \bibinfo{pages}{224401} (\bibinfo{year}{2001}).

\bibitem[{\citenamefont{Albuquerque et~al.}(2011)\citenamefont{Albuquerque,
  Schwandt, Hetenyi, Capponi, Mambrini, and Lauchli}}]{plaquette}
\bibinfo{author}{\bibfnamefont{A.~F.} \bibnamefont{Albuquerque}},
  \bibinfo{author}{\bibfnamefont{D.}~\bibnamefont{Schwandt}},
  \bibinfo{author}{\bibfnamefont{B.}~\bibnamefont{Hetenyi}},
  \bibinfo{author}{\bibfnamefont{S.}~\bibnamefont{Capponi}},
  \bibinfo{author}{\bibfnamefont{M.}~\bibnamefont{Mambrini}}, \bibnamefont{and}
  \bibinfo{author}{\bibfnamefont{A.~M.} \bibnamefont{Lauchli}}
  (\bibinfo{year}{2011}), \eprint{arXiv:1102.5325}.

\bibitem[{\citenamefont{Levin and Senthil}(2004)}]{levinsenthil}
\bibinfo{author}{\bibfnamefont{M.}~\bibnamefont{Levin}} \bibnamefont{and}
  \bibinfo{author}{\bibfnamefont{T.}~\bibnamefont{Senthil}},
  \bibinfo{journal}{Phys. Rev. B} \textbf{\bibinfo{volume}{70}},
  \bibinfo{pages}{220403} (\bibinfo{year}{2004}).

\bibitem[{\citenamefont{Calabrese
  et~al.}(2003{\natexlab{a}})\citenamefont{Calabrese, Pelissetto, and
  Vicari}}]{vicari2003}
\bibinfo{author}{\bibfnamefont{P.}~\bibnamefont{Calabrese}},
  \bibinfo{author}{\bibfnamefont{A.}~\bibnamefont{Pelissetto}},
  \bibnamefont{and} \bibinfo{author}{\bibfnamefont{E.}~\bibnamefont{Vicari}}
  (\bibinfo{year}{2003}{\natexlab{a}}), \eprint{cond-mat/0306273}.

\bibitem[{\citenamefont{Lang and Assaad}(unpublished)}]{assaad2010c}
\bibinfo{author}{\bibfnamefont{T.~C.} \bibnamefont{Lang}} \bibnamefont{and}
  \bibinfo{author}{\bibfnamefont{F.~F.} \bibnamefont{Assaad}}
  (\bibinfo{year}{unpublished}).

\bibitem[{\citenamefont{Fradkin et~al.}(2004)\citenamefont{Fradkin, Huse,
  Moessner, Oganesyan, and Sondhi}}]{fradkin2004}
\bibinfo{author}{\bibfnamefont{E.}~\bibnamefont{Fradkin}},
  \bibinfo{author}{\bibfnamefont{D.~A.} \bibnamefont{Huse}},
  \bibinfo{author}{\bibfnamefont{R.}~\bibnamefont{Moessner}},
  \bibinfo{author}{\bibfnamefont{V.}~\bibnamefont{Oganesyan}},
  \bibnamefont{and} \bibinfo{author}{\bibfnamefont{S.~L.}
  \bibnamefont{Sondhi}}, \textbf{\bibinfo{volume}{69}}, \bibinfo{pages}{224415}
  (\bibinfo{year}{2004}), \eprint{Phys. Rev. B}.

\bibitem[{\citenamefont{Polyakov}(1987)}]{polyakovbook}
\bibinfo{author}{\bibfnamefont{A.~M.} \bibnamefont{Polyakov}},
  \emph{\bibinfo{title}{Gauge Fields and Strings}} (\bibinfo{publisher}{Harwood
  Academic, New York}, \bibinfo{year}{1987}).

\bibitem[{\citenamefont{Ma and Berker}(1984)}]{ffisingcubic1}
\bibinfo{author}{\bibfnamefont{D.~B.~M.} \bibnamefont{Ma}} \bibnamefont{and}
  \bibinfo{author}{\bibfnamefont{A.~N.} \bibnamefont{Berker}},
  \bibinfo{journal}{Phys. Rev. B} \textbf{\bibinfo{volume}{30}},
  \bibinfo{pages}{1362} (\bibinfo{year}{1984}).

\bibitem[{\citenamefont{Jalabert and Sachdev}(1991)}]{ffisingcubic2}
\bibinfo{author}{\bibfnamefont{R.~A.} \bibnamefont{Jalabert}} \bibnamefont{and}
  \bibinfo{author}{\bibfnamefont{S.}~\bibnamefont{Sachdev}},
  \bibinfo{journal}{Phys. Rev. B} \textbf{\bibinfo{volume}{44}},
  \bibinfo{pages}{682} (\bibinfo{year}{1991}).

\bibitem[{\citenamefont{Calabrese
  et~al.}(2003{\natexlab{b}})\citenamefont{Calabrese, Pelissetto, and
  Vicari}}]{vicari2003b}
\bibinfo{author}{\bibfnamefont{P.}~\bibnamefont{Calabrese}},
  \bibinfo{author}{\bibfnamefont{A.}~\bibnamefont{Pelissetto}},
  \bibnamefont{and} \bibinfo{author}{\bibfnamefont{E.}~\bibnamefont{Vicari}},
  \bibinfo{journal}{Phys. Rev. B} \textbf{\bibinfo{volume}{67}},
  \bibinfo{pages}{054505} (\bibinfo{year}{2003}{\natexlab{b}}).

\bibitem[{\citenamefont{Sandvik}(2007)}]{sandvik2007}
\bibinfo{author}{\bibfnamefont{A.~W.} \bibnamefont{Sandvik}},
  \bibinfo{journal}{Phys. Rev. Lett} \textbf{\bibinfo{volume}{98}},
  \bibinfo{pages}{227202} (\bibinfo{year}{2007}).

\bibitem[{\citenamefont{Melko and Kaul}(2007)}]{kaul2007}
\bibinfo{author}{\bibfnamefont{R.~G.} \bibnamefont{Melko}} \bibnamefont{and}
  \bibinfo{author}{\bibfnamefont{R.~K.} \bibnamefont{Kaul}},
  \bibinfo{journal}{arXiv:0707.2961}  (\bibinfo{year}{2007}).

\bibitem[{\citenamefont{Balents et~al.}(2005)\citenamefont{Balents, Bartosch,
  Burkov, Sachdev, and Sengupta}}]{balentsvortex}
\bibinfo{author}{\bibfnamefont{L.}~\bibnamefont{Balents}},
  \bibinfo{author}{\bibfnamefont{L.}~\bibnamefont{Bartosch}},
  \bibinfo{author}{\bibfnamefont{A.}~\bibnamefont{Burkov}},
  \bibinfo{author}{\bibfnamefont{S.}~\bibnamefont{Sachdev}}, \bibnamefont{and}
  \bibinfo{author}{\bibfnamefont{K.}~\bibnamefont{Sengupta}},
  \bibinfo{journal}{Phys. Rev. B} \textbf{\bibinfo{volume}{71}},
  \bibinfo{pages}{144508} (\bibinfo{year}{2005}).

\bibitem[{\citenamefont{Burkov and Balents}(2005)}]{balentsburkov}
\bibinfo{author}{\bibfnamefont{A.~A.} \bibnamefont{Burkov}} \bibnamefont{and}
  \bibinfo{author}{\bibfnamefont{L.}~\bibnamefont{Balents}},
  \bibinfo{journal}{Phys. Rev. B} \textbf{\bibinfo{volume}{72}},
  \bibinfo{pages}{134502} (\bibinfo{year}{2005}).

\bibitem[{\citenamefont{Grover and Senthil}(2008)}]{senthil2007}
\bibinfo{author}{\bibfnamefont{T.}~\bibnamefont{Grover}} \bibnamefont{and}
  \bibinfo{author}{\bibfnamefont{T.}~\bibnamefont{Senthil}},
  \bibinfo{journal}{Phys. Rev. Lett.} \textbf{\bibinfo{volume}{100}},
  \bibinfo{pages}{156804} (\bibinfo{year}{2008}).

\bibitem[{\citenamefont{Abanov and Wiegmann}(2000)}]{abanov2000}
\bibinfo{author}{\bibfnamefont{A.~G.} \bibnamefont{Abanov}} \bibnamefont{and}
  \bibinfo{author}{\bibfnamefont{P.~B.} \bibnamefont{Wiegmann}},
  \bibinfo{journal}{Nucl. Phys. B} \textbf{\bibinfo{volume}{570}},
  \bibinfo{pages}{685} (\bibinfo{year}{2000}).

\bibitem[{\citenamefont{Abanov}(2001)}]{abanov2001}
\bibinfo{author}{\bibfnamefont{A.~G.} \bibnamefont{Abanov}},
  \bibinfo{journal}{JHEP} \textbf{\bibinfo{volume}{10}}, \bibinfo{pages}{030}
  (\bibinfo{year}{2001}).

\bibitem[{\citenamefont{Ryu et~al.}(2009)\citenamefont{Ryu, Mudry, Hou, and
  Chamon}}]{ryuwzw}
\bibinfo{author}{\bibfnamefont{S.}~\bibnamefont{Ryu}},
  \bibinfo{author}{\bibfnamefont{C.}~\bibnamefont{Mudry}},
  \bibinfo{author}{\bibfnamefont{C.-Y.} \bibnamefont{Hou}}, \bibnamefont{and}
  \bibinfo{author}{\bibfnamefont{C.}~\bibnamefont{Chamon}},
  \bibinfo{journal}{Phys. Rev. B} \textbf{\bibinfo{volume}{80}},
  \bibinfo{pages}{205319} (\bibinfo{year}{2009}).

\end{thebibliography}

\end{document}